# Title: Validation of apparent intra- and extra-myocellular lipid content indicator using spiral spectroscopic imaging at 3T

Antoine Naëgel[1,2], Magalie Viallon[1], Jabrane Karkouri[1,2,3], Thomas Troalen[2], Pierre Croisille[1], Hélène Ratiney[1],

[1] Université de Lyon, INSA-Lyon, Université Claude Bernard Lyon 1, UJM-Saint Etienne, CNRS, Inserm, CREATIS UMR 5220, U1206, Lyon, France, [2] Siemens Healthineers, France, [3] Wolfson Brain Imaging Center, University of Cambridge, Cambridge, United Kingdom,

(Character with space: 4987/5000)

**Introduction:** This work presents a fast and simple method based on spiral MRSI for mapping the IMCL and EMCL[1,2] apparent content, which is a challenging task[3,4] and it compares this indicator to classical quantification results in muscles of interest.

**Methods**: A spiral MRSI sequence was developed on a 3T clinical MRI (MAGNETOM PRISMA, Siemens Healthineers, Erlangen, Germany). Main parameters were TR/TE=2s/2ms, FOV=200x200x25mm, spatial resolution=64x64, voxel size=3.1x3.1x25mm, temporal resolution=500ms, temporal points=1024, spatial interleaving=22 and temporal interleaving=5, TAcq=3min48s. Spiral MRSI was performed using a dual resonance 1H/31P transmit/receive coil (Rapid GMBH, Würzburg, Germany) positioned under the right calf of 16 volunteers. A Fast (SE-EPI-based) diffusion-weighted and a T1 vibe Dixon sequence was subsequently acquired to derive the fibres' orientation, high-resolution water, fat and Fat Fraction (FF) images. MRSI data were analyzed using homemade processing tools (Matlab). After standard MRSI reconstruction, the analysis of the spectra was centered on the characteristic peaks of IMCL and EMCL (1.3ppm-1.5ppm). To realize an automatic phasing and frequency registration of the spectra (unsuppressed water), a Fourier transform on the absolute value of the time domain signal was performed[5]. Then, the evolution of the cumulative sum of the amplitudes (CSA) of a fixed area, defined between 1.1ppm and 1.7ppm, was used to analyze the apparent content of IMCL and EMCL for each voxel. Mapping the value of this curve index at 1.40 ppm enables displaying the apparent content of IMCL over EMCL in the fat component (Fig1). The new proposed apparent IMCL/EMCL content indicator was compared to the classical IMCL/(EMCL+IMCL) ratio quantified using LCModel fitting method (basis set "muscle-5") on region of Interested (ROI) selected in soleus medial (SM), and gastrocnemius medial (GM) muscles (Fig2). FF was calculated with the Dixon acquisition and compared to FF obtained by 2 methods derived from the MRSI data: one based on the CSA of the lipid and water signal, and another based on the quantification of the signal with LCModel.

**Results:** The average FF obtained by the 3 methods on the muscles of interest were resumed in Fig3 and coherent with previously obtained values in volunteers[6]. The apparent content indicator and its quantitative equivalent were both significantly different between the GM and SM muscles. There is a significant positive correlation between the apparent content indicator and its quantitative equivalent (Fig4). In addition, the GM muscle fibers displayed an overall alignment along the direction of the B0 field. In contrast, the SM's ones had an orientation between the Y and Z-axis.

**Discussion:** The MRS measure of IMCL is influenced by the quantity in the tissue and the orientation of the fibers. Our results are in agreement with literature[7]: The SM muscle had a high IMCL content due to its high percentage of type I fibers, with a pronounced angular fiber orientation to B0; the GM muscle had smaller IMCL content and a less pronounced angular fiber orientation. Despite a less advantageous fiber orientation, the observation of IMCL was feasible in the SM due to its high level in this muscle. The quantification of IMCL and EMCL on MRSI data is fastidious due to phase and frequency shifts from voxel to voxel and results in time-consuming data processing. The rapid analysis provided by the apparent indicator can be used to rapidly generate maps of the intra/extra lipid distribution. Also, this map enables analyzing lipid MR spectra as a function of fiber orientation, which is known to influence.

**Conclusion**: The proposed exploration technique is a promising, fast, and straightforward approach to map the apparent content of IMCL compared to EMCL lipids. It is correlated with results from standard quantification procedure and appear to be robust to signal-to-signal fluctuations related to B0 variations. Further work should evaluate the reproducibility prior to transfer to clinic for longitudinal studies. This preliminary work highlights the potential of this high-resolution spiral spectroscopic imaging technique to provide more insights on the coupling between structure, function, metabolism, energy consumption and the underlying pathophysiology in muscles, especially in the context of refining our understanding of impaired exercise performance, intolerance to sustained exercise and premature fatigability.

**Acknowledgements:** This work was partly supported by the LABEX PRIMES (ANR-11-LABX-0063), Siemens Healthineers and Jabrane Karkouri was supported by the European Union's Horizon 2020 research and innovation program under grant agreement No-801075.

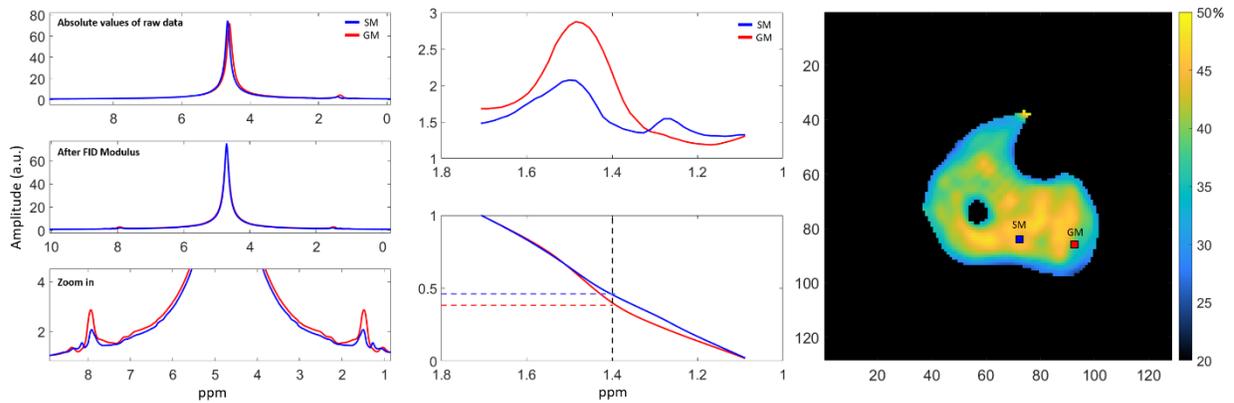

Figure 1: Apparent Content Indicator procedure. Left: FID modulus: automatic phasing and frequency registration. Middle: Magnitude spectrum centered on EMCL and IMCL and the evolution of the CSA. Right: apparent content indicator mapping and ROI highlighted for SM, GM muscles.

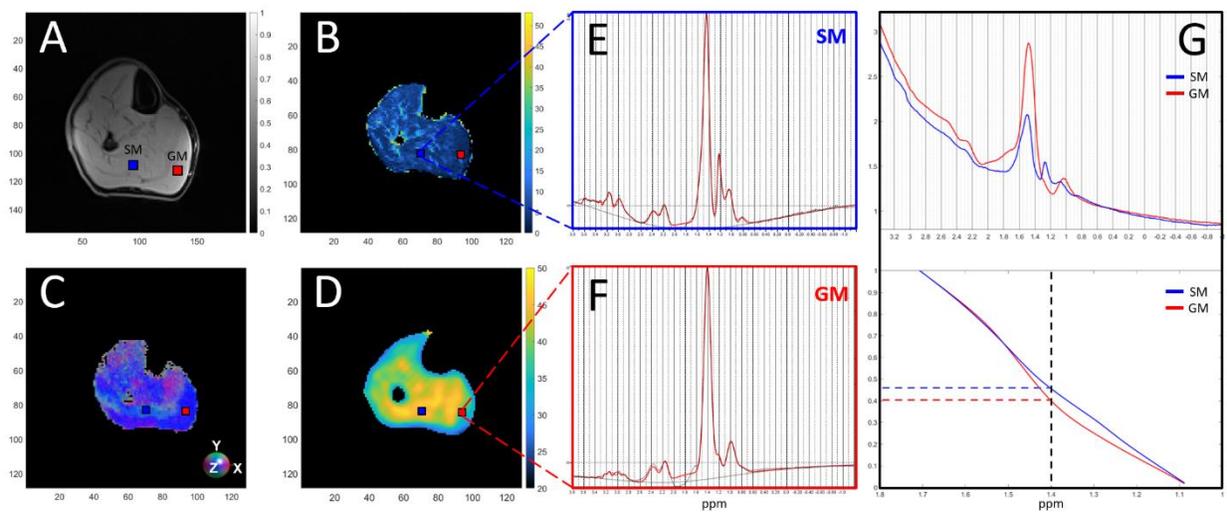

Figure 2: (A) T1-vibe Dixon water (B) Dixon FF map (C) fiber orientation as EigVector (X,Y,Z=R,G,B) (D) apparent content indicator map. Spectrum obtained in SM muscle (E) and GM muscle (F), processed with LCModel and (G) with the apparent content indicator procedure.

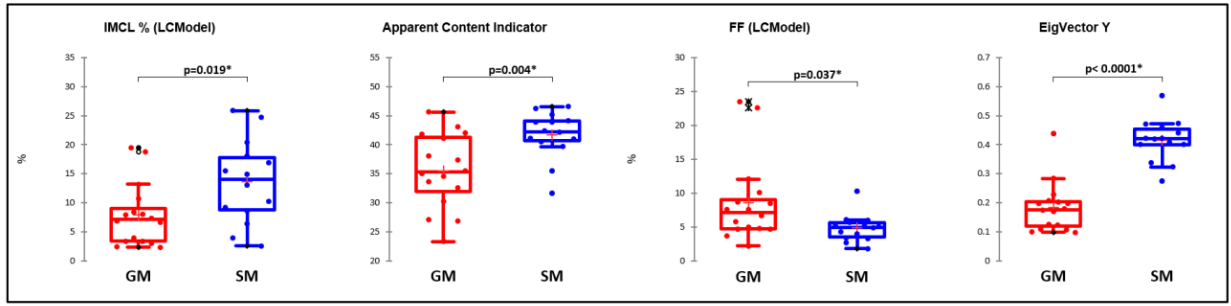

| | GM (N=16) | SM (N=14) | P-values* | Variation |
|---|---|---|---|---|
| IMCL % (LCModel) | 7.84 (5.37) | 13.59 (7.22) | **0.019** | ↗ |
| Apparent Content Indicator | 35.49 (6.43) | 41.69 (4.09) | **0.004** | ↗ |
| FF % (LCModel) | 8.6 (6.15) | 4.82 (2.01) | **0.037** | ↘ |
| FF (Indicator) | 9.16 (2.58) | 7.59 (0.85) | **0.039** | ↘ |
| FF % (Dixon) | 12.81 (7.51) | 10.71 (5.48) | 0.396 | |
| EigVector X | 0.23 (0.08) | 0.25 (0.09) | 0.521 | |
| EigVector Y | 0.18 (0.09) | 0.42 (0.07) | **0.000** | ↗ |
| EigVector Z | 0.92 (0.08) | 0.79 (0.07) | **0.000** | ↘ |

Figure 3: Table: Results of the statistical analysis of GM and SM muscles: mean (SD). *T-test with statistical significance set to $p<0.05$. Arrows provide the trend of the significant changes. Boxplots: Relevant parameter with significant difference between GM and SM muscles.

| Variables | IMCL % (LCModel) | Apparent Content Indicator | FF % (LCModel) | FF % (Indicator) | FF % (Dixon) | EigVector X | EigVector Y | EigVector Z |
|---|---|---|---|---|---|---|---|---|
| IMCL % (LCModel) | 1 | **0.693** | **-0.410** | -0.271 | 0.134 | 0.213 | **0.640** | **-0.574** |
| Apparent Content Indicator | **0.693** | 1 | **-0.709** | **-0.592** | -0.276 | 0.236 | **0.533** | **-0.510** |
| FF % (LCModel) | **-0.410** | **-0.709** | 1 | **0.938** | **0.600** | -0.006 | -0.187 | 0.139 |
| FF % (Indicator) | -0.271 | **-0.592** | **0.938** | 1 | **0.690** | 0.028 | -0.110 | 0.082 |
| FF % (Dixon) | 0.134 | -0.276 | **0.600** | **0.690** | 1 | 0.254 | 0.182 | -0.244 |
| EigVector X | 0.213 | 0.236 | -0.006 | 0.028 | 0.254 | 1 | 0.157 | **-0.633** |
| EigVector Y | **0.640** | **0.533** | -0.187 | -0.110 | 0.182 | 0.157 | 1 | **-0.834** |
| EigVector Z | **-0.574** | **-0.510** | 0.139 | 0.082 | -0.244 | **-0.633** | **-0.834** | 1 |

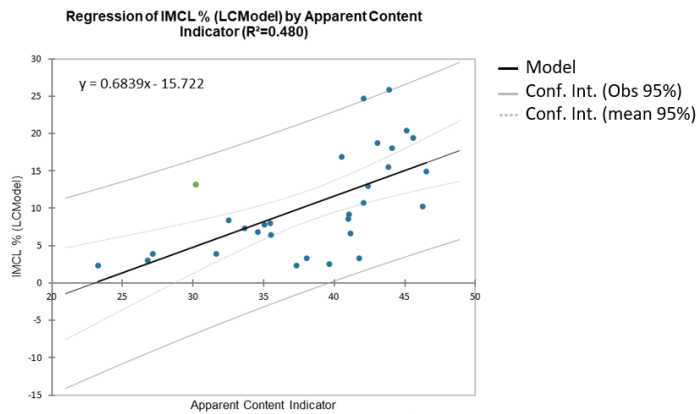

Figure 4: Table: Spearman correlation matrix, correlation coefficients in bold have a $p<0.05$. Graph: Linear regression between the IMCL % and the apparent content indicator.